\documentstyle[12pt]{article}
\begin{document}
\baselineskip=22pt plus 0.2pt minus 0.2pt
\lineskip=22pt plus 0.2pt minus 0.2pt

\begin{center}
\large
Modelling the Off-Shell Dependence\\
of $\pi^{o}- \eta$ Mixing with Quark Loops\\

\vspace{0.50in}

Kim Maltman

\vspace{0.10in}

Department of Mathematics and Statistics, York University\\
4700 Keele Street,\\
North York, Ontario, CANADA M3J 1P3\\

\vspace{0.10in}

and\\

\vspace{0.10in}

Physics Division, Los Alamos National Laboratory\\
Los Alamos, New Mexico  87545  USA\\

\vspace*{0.50in}

and\\

\vspace{0.50in}

T.\ Goldman\\

\vspace{0.10in}

Theoretical Division, Los Alamos National Laboratory\\
Los Alamos, New Mexico 87545  USA

\vspace{1.00in}

ABSTRACT
\end{center}

It is shown that including form factors for the quark-pseudoscalar
meson couplings of the Georgi-Manohar chiral quark model allows one to
obtain the leading off-shell dependence of $\pi^{o}- \eta$ mixing (as
predicted by chiral perturbation theory) from the effect of quark loops
on the meson propagators.  Implications for $\rho^{o} - \omega$ mixing
and for the effects on meson mixing contributions to few body charge
symmetry breaking observables are also discussed.

\pagebreak

\noindent INTRODUCTION

\vspace{0.10in}

It is commonly held that the bulk of non-electromagnetic charge
symmetry breaking (CSB) in few-nucleon systems is due to
isoscalar-isovector mixing in the intermediate meson propagators of
one-boson-exchange graphs.  The dominant contributions of this type are
associated with $\rho^{o} - \omega$ and, to a lesser extent, $\pi^{o} -
\eta$ mixing [1-4].  Recently, however, it has been pointed out that
some problems exist with the standard evaluations of these
contributions.

First, recall that the $\rho^{o} - \omega$ mixing matrix element is
obtained from experimental data on $e^{+} e^{-} \rightarrow \pi^{+}
\pi^{-}$ in the $\rho - \omega$ interference region and therefore
corresponds to $q^{2} \simeq m_{\omega}^{2}$.  The standard analysis
assumes this matrix element is $q^{2}$-independent and, therefore, uses
the experimental value unchanged in the NN scattering region, for which
$q^{2} < 0$.  Not only has the validity of this \underline{ansatz}
never been tested, but, in Ref.\ 5, using a model in which $\rho^{o} -
\omega$ mixing is generated by an intermediate quark loop as a
consequence of the inequality of up and down quark masses, Goldman,
Henderson, and Thomas (GHT) raise the possibility of significant
$q^{2}$-dependence of the $\rho^{o} - \omega$ mixing matrix element.

Second, a recent evaluation [6] of $\pi^{o} - \eta$ mixing to one
loop in chiral perturbation theory (ChPT), shows that 1) the mixing
matrix element is, indeed, $q^{2}$-dependent (varying by $\sim$ 20\%
over a range from $q^{2}\simeq m^2_{\eta}$ to $q^2 \simeq -
m^2_{\eta}$, comparable to that involved in the extrapolation of the
$\rho^{o} - \omega$ matrix element from $q^{2} \simeq m_{\omega}^{2}$
to the scattering region) and 2) the magnitude of the mixing, even
without this $q^{2}$-dependence, is less than obtained by the model
analysis of Ref.\ 3c.  The results of Ref.\ 6, being at one loop in
ChPT, provide only the linear-in-$q^{2}$ dependence of $\pi^{o} -
\eta$ mixing; the higher order dependence is unconstrained but will
be small in the region of validity of the chiral expansion ($|q^{2}|
\stackrel{<}{\sim} m_{\eta}^{2}$).  Since the convergence of the
chiral expansion to one loop in this region has been extensively
tested [7,8], and the framework of ChPT is a rigorous consequence of
QCD, the results of Ref.\ 6 should be considered very reliable (up to
an overall $\sim$20\% scale uncertainty associated with the treatment
of electromagnetic corrections of pseudoscalar masses only to leading
order in the chiral expansion [6,9]).

The aim of the present paper is to use the constraints of Ref.\ 6 to
test the type of model building which underlies the GHT approach.  As
we will see, the modelling does, indeed, succeed in reproducing the
correct $q^{2}$-dependence of $\pi^{o} - \eta$ mixing, which
considerably strengthens the case of GHT that, as a consequence of
the $q^{2}$-dependence of the mixing, the dominant $\rho^{o} -
\omega$ mixing contribution to few-body CSB may be significantly
different than previously thought.\\

\noindent THE $Q^{2}$-DEPENDENCE of $\pi^{o} - \eta$ MIXING AND THE
QUARK LOOP MODEL

\vspace{0.10in}

Let us begin with a few general remarks.  These are necessitated by
the, apparently widely held, view that mixing matrix elements of the
sort we are discussing are expected to be ``naturally''
$q^{2}$-independent.  This is simply not the case.

In fact the whole assumption underlying the meson exchange framework is
that, at low energies, QCD reduces to an effective low energy theory
involving only composite hadronic fields (meson, nucleons, deltas...).
Let us assume, for the sake of the argument in this paper, that this
assumption is essentially correct. Then, whatever the effective theory
governing the interactions of these hadrons \underline{is}, we know
that, as an effective low energy theory, it will be described by a
non-renormalizable Lagrangian in which all terms not explicitly
forbidden by the symmetries of the underlying theory (QCD) occur. In
particular, there will be terms in the meson sector of the theory which
involve the quark mass matrix and higher powers of derivatives, which
will naturally lead to $q^{2}$-dependent meson mixing. The pseudoscalar
sector, where we actually know something about the effective theory
beyond leading order in the momentum expansion, is one explicit example
of this general principle. The standard \underline{ansatz}, of taking,
eg., the $\rho^{o} - \omega$ matrix element to be independent of
$q^{2}$, is thus incompatible with the assumptions underlying the
framework in which such mixing is to be used to generate CSB in
few-body systems. Of course, without knowing the terms of effective
Lagrangian involving the vector mesons beyond leading order, one does
not know what the magnitude of the $q^{2}$-dependence is, and hence
whether the effect of the $q^{2}$-running is important or not. Given
the pseudoscalar result and the natural (QCD) scales involved, however,
it is unlikely that the effect will be negligible. The fact that the
vector mesons are much less point-like than the pseudscalar mesons
further supports this contention. \footnote {Note also that the
non-renormalizable structure of the effective low energy Lagrangian
will\\
lead to a $q^{2}$-dependence of the ``renormalized'' (in the sense of
the quark mass/momentum\\
expansion) self-energies for the composite fields.  This means that
the meson propagators in the\\
effective theory may differ significantly from the form $$ \Delta(q)
= i/(q^{2} - m^{2}) $$ away from the pole (i.\ e.\ especially in the
spacelike region relevant to NN scattering in meson\\
exchange models). The pseudoscalar mesons are special in this regard
since, as one can readily\\
see from the structure the effective Lagrangian [see e.\ g.\ Ref.\
7], the $q^{2}$-dependence of the\\
renormalized self-energy enters first at two-loop order and hence
will be very small, by the\\
usual power counting arguments, in the region of validity of the
chiral expansion.  This is a\\
special property associated with the chiral constraints of the
pseudoscalar sector and will not be\\
a property, e.\ g.\ of vector meson fields.}

Let us now turn to the question at hand, namely, whether or not a
quark-loop model of the GHT type is capable of reproducing the
behavior of $\pi^{o} - \eta$ mixing known from ChPT.  To one-loop in
ChPT, Ref.\ 6 shows that one obtains a $q^{2}$-dependent $\pi_{3} -
\pi_{8}$ mixing angle given by
$$
\theta(q^2) = \frac{\sqrt{3}(m^2_{K^o} - m^2_{K^+})_{QCD}}{(m^2_K -
m^2_{\pi})} $$
$$
\left[ 1 + \Delta_{GMO} + \frac{1}{16 \pi^2 f^2}
\left(\frac{m^2_{\eta}}{m^2_{\eta} - m^2_{\pi}}\right)
\right. 
$$
$$
\times \left( 3m^2_{\eta} \ln (m^2_K / m^2_{\eta}) + m^2_{\pi} \ln
(m^2_K /
m^2_{\pi}) \right)
$$
$$
\left. 
+ \left( \frac{q^2 + m^2_{\eta}}{32\pi^2f^2}\right) \left( 1 +
\left(\frac{m^2_{\pi}}{m^2_K - m^2_{\pi}}\right)
\ln( m^2_{\pi} / m^2_K) \right) \right]
$$
$$
\eqno (1)
$$

\noindent where $\pi_{3}, \pi_{8}$ are the unmixed states to which
the physical $\pi^{o},\eta$ states reduce in the limit of isospin
symmetry, $\Delta_{GMO}$ is the Gell-Mann-Okubo discrepancy
$$
\Delta_{GMO} = (4m^{2}_{K} - m^{2}_{\pi} -
3m^{2}_{\eta})/(m^{2}_{\eta} - m^{2}_{\pi}), \eqno (2)
$$
$f$ is a parameter of the chiral expansion (equal to $f_{\pi}$ in
leading order) and\\
$(m^{2}_{K^{o}} - m^{2}_{K^{+}})_{QCD}$ is the contribution to the kaon
mass splitting due to $m_{d} \neq m_{u}$, obtained by correcting the
observed splitting for electromagnetic effects. This correction is
usually performed using only leading order results in the
electromagnetic chiral expansion (Dashen's theorem [10]). If this
procedure were accurate, the corrections to (1), which enter only at
6th order in the low-energy expansion, would be small in the region of
small $q^{2}$, say $|q^{2}| \stackrel{<}{\sim} m^{2}_{\eta}$.

An old attempt by Socolow [11] to saturate the Cottingham formula for
the kaons with K, K$^{*}$ states, however, indicates that there may
be significant corrections to Dashen's theorem. This possibility is
also discussed in Ref.\ 9, on the basis of the size of the relevant
electromagnetic chiral logarithms. At present, both the kaon
electromagnetic splitting obtained from Dashen's theorem and the
larger Socolow value are compatible with other isospin breaking data
[9]. The larger Socolow value would lead to a value of
$\;(m^{2}_{K^{o}} - m^{2}_{K^{+}})_{QCD}\,$ 30\% \underline{larger}
than that obtained using Dashen's theorem. There is thus an overall
scale uncertainty in (1) (which uncertainty, however, can be reduced
by improved experimental results,\\
e.\ g.\ on isospin breaking in $K_{e3}$).

Our aim now is to see whether a quark-loop model of the GHT type
succeeds in reproducing the constraints just discussed. Two assumptions
underlie the GHT approach: first, that the behavior of the full
low-energy meson theory can be obtained by integrating out quarks from
a theory consisting of free mesons coupled to quarks and, second, that
the effect of the quark loops can be adequately represented by keeping
only those loops generated by the lowest order quark-meson couplings of
the effective quark-meson theory using a) a monopole form factor at
each quark-meson vertex, and b) free constituent quark propagators. The
basic philosophy behind these assumptions is that the higher order
terms in the effective meson Lagrangian arise from the non-pointlike,
quark substructure of the mesons, and that incorporating this
substructure in a way that (through the monopole form factor parameter,
$\Lambda$) is capable of reflecting the actual meson size should allow
one to reproduce the behavior associated with these terms.

For the problem at hand we, therefore, start with a model which
incorporates both quarks and pseudoscalar mesons.  The model must,
moreover, properly respect the approximate chiral symmetry with which
the pseudoscalars are associated as pseudo-Goldstone bosons.  A
natural choice is the Georgi-Manohar chiral quark model [12,13].  The
model describes an effective theory of massive constituent $u,d,s$
quarks and the pseudoscalar octet, \{$\pi^{a}$\}, transforming
non-linearly under chiral $SU(3)_{L} \times SU(3)_{R}$ according to

$$
\xi \rightarrow \xi^{\prime} = L \xi U^{+} = U \xi R^{+} \eqno (3a)
$$
$$
q \rightarrow q^{\prime} = Uq \eqno (3b)
$$
where $\xi = exp(i {\bf \lambda} \cdot {\bf \pi}/2f)$ is the square
root of the matrix $\Sigma = exp(i {\bf \lambda} \cdot {\bf
\pi}/f)\:$,$ \: $ \{ $\lambda^{a}$ \} are the usual Gell-Mann
matrices and $f$ is as in Eqn.\ (1) above. $\Sigma$ transforms
linearly under $SU(3)_{L}\, \times \, SU(3)_{R}$,
$$
\Sigma \rightarrow \Sigma^{\prime} = L\,\Sigma\,R^{+}. \eqno (4)
$$
In (3) and (4), $U$ reduces to the usual flavor $SU(3)$ matrix for an
$SU(3)_{V}$ transformation and is a non-linear function of L, R,
\{$\pi^{a}$\} otherwise.  In the chiral limit (in which the model is
to be invariant under the transformations (3a),(3b)), if we consider
only terms with zero or one derivative (the leading terms in the
momentum expansion), the effective Lagrangian of the model is
$$
\L^{(o)}_{eff} = -m \bar{q} q + i \bar{q} \, D \hspace{-.12in}/ \, \,
q + \, g_{A} \,  \bar{q} \, {\bf A}\hspace{-.10in}/ \, \gamma \, _{5}
\, q \eqno (5)
$$
where the covariant derivative, $D_{\mu}$, is given by

$$
D_{\mu} = \partial_{\mu} - i {\bf V}_{\mu} \eqno (6)
$$
and the vector and axial vector fields ${\bf V}_{\mu}$, ${\bf
A}_{\mu}$ are defined by
$$
{\bf V}_{\mu} = \frac{1}{2} (\xi^{+} \partial_{\mu} \xi + \xi
\partial_{\mu} \xi^{+} )
$$
$$
{\bf A}_{\mu} =\frac{i}{2}(\xi^{+} \partial_{\mu} \xi - \xi
\partial_{\mu} \xi^{+}) \eqno (7)
$$

In Eqn.\ (5), $g_{A}$ is the constituent quark axial coupling, $g_{A}
\simeq 0.75$, and $m$ is the hypothetical constituent quark mass, in
the chiral limit.  At this stage chiral symmetry is unbroken, the
$\pi^{a}$ are all massless, isospin is exact, and there is no
$\pi_{3} - \pi_{8}$ mixing.  Chiral symmetry breaking is then
incorporated using the standard techniques of effective chiral
Lagrangians.  The leading symmetry breaking term involving both
quarks and pseudoscalars is
$$
\L_{B} = \frac{-Cm}{\Lambda_{\chi} ^{2}} \bar{q} (\xi \mu M \xi +
\xi^{+} \mu M \xi^{+}) q \eqno (8)
$$
where $\mu$ is a mass scale related to the quark condensate, $M$ is
the current quark mass matrix, $M = diag(m^{c}_{u}, m^{c}_{d},
m^{c}_{s}),$ $\Lambda_{\chi}$ is a chiral symmetry breaking scale
$\sim 1$GeV and $C$ is expected to be of order 1.  $C$ is in fact
constrained by the observation that the terms zeroth order in
$\pi^{a}$ in Eqn.\ (8) produce a splitting of the constituent $s$
mass from the constituent $u,d$ masses
$$
\delta m^{con}_s \equiv m^{con}_{s} - m^{con}_{u,d} \simeq \frac{2Cm
m_{K} ^{2}}{\Lambda_{\chi} ^{2}} \sim 200 MeV \eqno (9)
$$
where the lowest order mass relations for the pseudoscalars have been
used to set $\mu m_{s} \simeq m_{K} ^{2}$.  For further details of
the model, the reader is referred to Refs.\ 12,13.

According to the GHT ansatz, what we are now supposed to do is 1)
drop all additional terms in the chiral expansion for the quark,
pseudoscalar, and quark-pseudoscalar sectors, and 2) hope to generate
the effects of these terms by including a monopole form factor, $F
(k^{2}) = \Lambda^{2} / (\Lambda^{2} - k^{2})$, for each meson leg
coupling to a quark line (with $k$ the four-momentum flowing through
the vertex on the quark line) and considering the effect of quark
loops with free, constituent quark propagators.

Such loops contribute to the pseudoscalar propagators in two ways.
First, the axial vector coupling term in Eqn.\ (5) generates
pseudovector quark-pseudoscalar meson couplings, which in turn
generate two-vertex loops as in Fig.\ 1.  (Because of the
pseudovector nature of the couplings, these loops are proportional to
$q^{2}$ near $q^{2} = 0$ and do not contribute to the pseudoscalar
masses.)  Second, the terms of $\L_{B}$ second order in the
pseudoscalar fields produce tadpole diagrams, as in Fig.\ 2.  These
loop contributions are proportional to \underline{current} quark
masses and naturally lead to the usual relations between the
pseudoscalar squared-masses and the current quark masses,
$$
m^{2}_{\pi} = \mu (m^{c}_{u} + m^{c}_{d})
$$
$$
m^{2}_{K^+} = \mu(m^{c}_{s} + m^{c}_{u})
$$
$$
m^{2}_{K^o} = \mu(m^{c}_{s} + m^{c}_{d})
$$
$$
m^{2}_{\eta} = \mu(\frac{4}{3} m^{c}_{s} + \frac{1}{3} m^{c}_{u}+
\frac{1}{3} m^{c}_{d}) \eqno (10)
$$
up to an overall factor which is $\Lambda$-dependent, as we will see
in more detail below.

Requiring that we reproduce the correct leading mass relations would,
therefore, fix the value of $\Lambda$ to be used in the GHT ansatz.
Choosing $\Lambda$ in this way then automatically also fixes the
$\pi_{3} - \pi_{8}$ mixing angle to its correct, leading order value
$$
\theta^{leading} = \frac{\mu(m_{d} - m_{u})}{\sqrt{3}(m^{2}_{\eta} -
m^{2}_{\pi})}. \eqno (11)
$$
The real test of the GHT ansatz is then whether or not it is able to
reproduce the correct $q^{2}$-dependence of $\theta$, as given in
Eqn.\ (1).

To evaluate $\theta(q^2)$ in the model, we include the quark loops of
Figs.\ 1,2.  The $\pi_{3} - \pi_{8}$ inverse propagator then takes
the form

\renewcommand{\arraystretch}{2.0}
$$
{\bf \Delta}^{-1}(q^{2}) = \left[ \begin{array}{c}
{q^{2} - \pi_{33}(q^{2})}   \\  {- \pi_{38}(q^{2})} \end{array} \\
\hspace{0.3in}
\begin{array}{c}
{- \pi_{38}(q^{2})}         \\      {q^{2}- \pi_{88}(q^{2})}
\end{array}
\right]. \eqno (12)
$$

In Eqn.\ (11), $\pi_{33}, \pi_{38}, \pi_{88}$ all include both
tadpole and two-vertex-loop contributions.  The former are
$q^{2}$-independent, the latter $q^{2}$-dependent (and beginning at
$O(q^{2})$ near $q^{2} = 0$).  $\pi_{38}$, moreover, is proportional
to $(m^{c}_{d} - m^{c}_{u})$.  For small $q^{2}$ (the only values for
which we have a constraint) we may write
$$
\pi_{kl} (q^{2}) = \pi^{(0)}_{kl} + q^{2} \pi^{(1)}_{kl} \eqno (13)
$$
where the $\pi^{(0)}_{kl}$ are associated with the tadpoles and the
$\pi^{(1)}_{kl}$ with the two-vertex loops.  Then the angle which
diagonalizes ${\bf \Delta}^{-1}$, to $O(m_{d} - m_{u})$ (the same
order as the ChPT result in Eqn.\ (1)), is given by
$$
\theta(q^{2}) = \frac{- \pi_{38} (q^{2})}{(\pi_{88} (q^{2}) -
\pi_{33} (q^{2}))}
$$
$$
\simeq \frac{- \pi^{(0)}_{38}}{\pi^{(0)}_{88} - \pi^{(0)}_{33}}
\left[ 1 + q^{2} \left(\frac{\pi^{(1)}_{38}}{\pi^{(0)}_{38}} -
\frac{(\pi^{(1)}_{88} - \pi^{(1)}_{33})}{(\pi^{(0)}_{88} -
\pi^{(0)}_{33})} \right) \right]
\eqno (14)
$$
where we have kept only terms up to $O(q^{2})$ in the expansion
occuring on the second line of Eqn.\ (14).  The appearance of the
leading order expression as an overall factor in Eqn.\ (14) ensures
that the scale of $\theta(q^{2})$ at $q^{2} = 0$ is the leading order
result,\\
Eqn.\ (1), when $\Lambda$ is chosen so as to give the leading order
results for the pseudoscalar masses.  There are, in fact, corrections
to the leading order pseudoscalar mass expressions at one-loop in ChPT
which would affect the choice of $\Lambda$, but these are not fixed
numerically with great precision because of residual uncertainties in
the higher order low-energy constants of Ref.\ 7.  The resulting
overall scale uncertainty in Eqn.\ (14) is not, however, a practical
difficulty for three reasons.  First, we know that the value should be
close to that required to give the correct leading order mass formula
and that this value of $\Lambda$ gives the leading order result for
$\theta$, which will be close to the correct magnitude at $q^{2} = 0$
even after one-loop corrections.  Second, there is a range of
uncertainty of 30\% in the overall scale of the result Eqn.\ (1) as a
consequence of the uncertainties in the electromagnetic corrections to
the kaon splitting.  And third, it turns out that the slope with
$q^{2}$ of the factor in brackets in Eqn.\ (14) is rather weakly
dependent on $\Lambda$.  We will, therefore, quote all results below
with $\Lambda$ fixed so as to give the leading order pseudoscalar mass
relations.  The potential error in the slope with $q^{2}$ is $\sim$
10\%.\\

\vspace{0.10in}

\noindent THE QUARK LOOP CONTRIBUTIONS

\vspace{0.10in}

To generate the loop contributions of Figs.\ 1,2 we require the
relevant quark-pseudoscalar meson couplings.  The pseudovector
couplings required for Fig.\ 1 come from the third term in Eqn.\ (5)
and are given by
$$
\frac{-g_{A}}{2f} \left(\bar{u} \gamma_{\mu} \gamma_{5} u - \bar{d}
\gamma_{\mu} \gamma_{5} d \right) \partial^{\mu} \pi_{3} -
\frac{g_{A}}{2 \sqrt{3} f} \left( \bar{u} \gamma_{\mu} \gamma_{5} u +
\bar{d} \gamma_{\mu} \gamma_{5} d - 2 \bar{s} \gamma_{\mu} \gamma_{5}
s \right) \partial^{\mu} \pi_{8}
$$
$$
\equiv ^{\hspace{0.10in} \sum}_{\stackrel{q = u,d,s}{a = 3,8}}
\left(\frac{g^{a,q}_{PV}}{f} \right) \bar{q} \gamma_{\mu} \gamma_{5}
q \partial^{\mu} \pi^{a} \eqno (15)
$$
where the second line defines the couplings, $g^{a,q}_{PV}$.  The
tadpoles of Fig.\ 2 are produced by terms which couple two
pseudoscalars to a quark line.  The contributions associated with the
vector current part of the second term in Eqn.\ (5) vanish by Lorentz
invariance.  The surviving tadpole contributions are then generated
by those pieces of $\L_{B}$ (Eqn.\ (8)) second order in the
\{$\pi^{a}$\}, the terms involving $\pi_{3}$, $\pi_{8}$ being

$$
\frac{c m \mu}{\Lambda^{2}_{\chi} f^{2}}
\left[ (m^{c}_{u} \bar{u} u + m^{c}_{d} \bar{d} d) (\pi_3)^2 +
\frac{2}{\sqrt{3}}\,(m^{c}_{u} \bar{u} u - m^{c}_{d} \bar{d} d)
\pi_{3} \pi_{8}\right.
$$
$$
\left.
+ \frac{1}{3}\,(4m^{c}_{s} \bar{s} s + m^{c}_{u} \bar{u} u +
m^{c}_{d} \bar{d} d) (\pi_8)^2  \right]
$$
$$
\equiv ^{\hspace{0.23in} {\sum}}_{\stackrel{q = u,d,s}{ab = 33,88,
38}} \; g^{ab,q}_{tad} m^{c}_{q} \bar{q} q \pi^{a} \pi^{b} \eqno (16)
$$
where the second line defines $g^{ab,q}_{tad}$.\\

{}From Eqn.\ (15) we find that the 2-vertex-loop contribution to the
$ab$ element of the mass matrix, $\pi^{(ab)}_{loop} (q^{2})$, is
given by
$$
\pi^{ab}_{loop} (q^{2}) = \sum_{q = u,d,s} \left( \frac{-i
g^{a,q}_{PV} g^{b,q}_{PV}}{2! \, \, f^{2}} q^{\mu} q^{\nu} \right)
$$
$$
\cdot \int \frac{d^{4}k}{(2\pi)^{4}} \frac{Tr[\gamma_{\mu}
\gamma_{5}(\frac{1}{2}q\hspace{-0.07in}/ + k\hspace{-0.08in}/ +
m_{q}) \gamma_{\nu} \gamma_{5} (-\frac{1}{2}q\hspace{-0.07in}/ +
k\hspace{-0.08in}/ + m_{q})]}{[(\frac{1}{2} q + k)^{2} -
m_q^{2}][(-\frac{1}{2} q + k)^{2} - m _q^{2}]}
\left(\frac{\Lambda^{2}}{(\Lambda^{2} - k^{2})} \right)^{2}. \eqno
(17)
$$

\vspace{0.20in}

\noindent This integral is readily evaluated using standard Feynman
parameter techniques.  The resulting expression is somewhat lengthy
and will not be quoted here since we require only the leading,
$O(q^{2})$, contribution.  This may be obtained straightforwardly
from the full expression, and is given by
$$
\pi^{(1)}_{ab} \equiv \frac{\partial \pi^{(ab)}_{loop}
(q^{2})}{\partial q^{2}} \Big{|}_{q^{2} = 0} = \sum_{q = uds}
\left(\frac{g^{a,q}_{PV} g^{b,q}_{PV}}{f^{2}} \right)
\frac{\Lambda^{4}}{8\pi^{2}}
$$
$$
\left[\frac{(\Lambda^{2} + 5m^{2}_{q})}{2(\Lambda^{2} -
m^{2}_{q})^{2}} - \frac{(2 \Lambda^{2} +
m^{2}_{q})m^{2}_{q}}{(\Lambda^{2} - m^{2}_{q})^{3}} ln \left(
\Lambda^{2}/m^{2}_{q} \right) \right]. \eqno (18)
$$

Similarly, from Eqn.\ (16), we obtain the ($q^{2}$-independent)
tadpole contribution to the mass matrix, $\pi_{ab}$
$$
\pi^{(0)}_{ab} = \; ^{\hspace{0.10in}\sum}_{q = uds} \;
{g^{ab,q}_{tad}} \; m^{c}_{q} (1 + \delta_{ab}) \int
\frac{d^{4}k}{(2\pi)^{4}} Tr \left[\frac{i}{k\hspace{-0.07in}/ -
m_{q}} \right] \left(\frac{\Lambda^{2}}{\Lambda^{2} - k^{2}}
\right)^{2} \eqno (19)
$$
$$
= (1 + \delta_{ab}) \; ^{\hspace{0.10in}\sum}_{q = uds} \;
g^{ab,q}_{tad}\; m^{c}_{q} \; Q \, (\Lambda, m)
$$
where
$$
Q(\Lambda,m) \equiv \frac{m\Lambda^{4}}{4 \pi^{2}}
\left[\frac{1}{(\Lambda^{2} - m^{2})} - \frac{m^{2}}{(\Lambda^{2} -
m^{2})^{2}} ln \left(\Lambda^{2}/m^{2} \right) \right] \eqno (20)
$$
and, in the second line of Eqn.\ (19), we have retained only those
terms linear in the current quark masses, $m^{c}_{q}$.  Eqn.\ (19),
together with Eqn.\ (9), implies
$$
\pi^{(0)}_{33} = \left[\frac{\delta m^{con}_{s} Q
(\Lambda,m)}{m^{2}_{K} f^{2}} \right] \mu (m^{c}_{u} + m^{c}_{d})
$$
$$
\pi^{(0)}_{88} = \left[\frac{\delta m^{con}_{s} Q(\Lambda,
m)}{m^{2}_{K} f^{2}} \right] \mu \left(\frac{4}{3} m^{c}_{s} +
\frac{1}{3} m^{c}_{u} + \frac{1}{3} m^{c}_{d} \right) \eqno (21)
$$
$$
\pi^{(0)}_{38} = - \left[\frac{\delta m^{con}_{s} Q (\Lambda,
m)}{m^{2}_{K} f^{2}} \right] \frac{\mu(m_{d} - m_{u})}{\sqrt{3}}
$$

\vspace{0.20in}

\noindent which, as promised, are the correct leading order results
providing $\Lambda$ is chosen so that

$$
Q(\Lambda, m) = m^{2}_{K} f^{2}/\delta m^{con}_{s}. \eqno (22)
$$

\vspace{0.10in}

\noindent RESULTS

\vspace{0.10in}

In order to employ the results of the last section in Eqn.\ (14), we
require a value for the constituent quark mass, $m$.  An extensive
analysis of the meson [14] and baryon [15] sectors, which includes
kinematic relativistic corrections, obtains $m = 220 MeV$, $\delta
m^{con}_{s} = 200 MeV$.  As an alternate value for $m$, and in order
to display the sensitivity to the parameters, we will also quote
results for $m = \frac{3}{2} m_{\pi^{o}}$, which would put the
threshold for the mixing matrix element in the correct location.
Rewriting Eqn.\ (1) as
$$
\theta^{ChPT}(q^{2}) = \theta(q^{2} = 0) [1 + c_1 q^{2}] \eqno (23)
$$
we have, from Eqn.\ (1), that $c_{1} = 0.279 GeV^{-2}$.

In Table 1 we present the model results for $c_{1}$.  The errors
quoted correspond to allowing $\Lambda$ to vary over a range for
which the coefficients of the leading order results in Eqn.\ (21)
vary by $\pm$20\% about the value 1 (a typical variation associated
with one-loop corrections in ChPT).  As may be seen from the Table,
this error amounts to $\sim$10\%.  $c_{1}$ also varies by $\sim$5\%
between $m = \frac{3}{2} m_{\pi^{o}}$ and $m = 220 MeV$.  A similar
variation is produced by varying $\delta m^{con}_{s}$ by $\sim$10\%.
As we see from the Table, the agreement with the ChPT result is quite
good.  This gives us confidence that the GHT type of modelling may
give reliable results for the off-shell dependence of mixing, at the
very least to lowest non-trivial order in $q^{2}$.

It should be pointed out, in passing, that the GHT calculation only
investigated the $q^{2}$-dependence of the off-diagonal $\rho -
\omega$ matrix element, $\pi_{\rho \omega}$, in the inverse
propagator, analogous to $\pi_{38}$ of Eqn.\ (11).  When one,
however, extracts the $\rho - \omega$ element of the propagator
itself and writes it in the form
$$
\frac{i \pi^{phen}_{\rho \omega} (q^{2})}{(p^{2} -
m^{2}_{\rho})(p^{2} - m^{2}_{\omega})}, \eqno (24)
$$
the $q^{2}$-dependence of $\pi^{phen}_{\rho \omega} (q^{2})$ arises
not only from the proportionality to $\pi_{\rho \omega}$, but also
from the $q^{2}$-dependence of the $\rho, \omega$ renormalized
self-energies.  The actual GHT predictions correspond to $\pi_{\rho
\omega}$ and \underline{not} $\pi^{phen}_{\rho \omega}$ and so should
not be employed, in their present form, in few-body CSB calculations.

The present calculation, however, suggests that the full version of
the GHT calculation for $\rho \omega$ mixing should provide a
reliable estimate of the off-shell behavior of the $\rho \omega$
matrix element, at least in the region of small $q^{2}$.  If one,
further, wished to take the success of the calculation as evidence
for the validity of the physical argument underlying the GHT ansatz,
one might hope that the higher order $q^{2}$-dependence would also be
well-modelled.  We have, of course, no way of demonstrating that this
will be the case; however, from both the present results, those of
ChPT, and from general principles, it is clear that a model which
passes the test of satisfying at least one known constraint should be
considered more reliable than one (the standard ansatz of a
$q^{2}$-independent matrix element) which fails it.  This means that
the few-body CSB contributions to such observables as the difference
of $n$ and $p$ scattering lengths and the non-Coulombic $A = 3$
binding energy difference need re-evaluation.  This can only be done
reliably (i.\ e.\ in the regime where the modelling has been tested)
to the extent that meson-mixing contributions are associated
primarily with the region of $q^{2}$ for which the leading
$q^{2}$-dependence of the mixing matrix element is dominant.

\noindent ACKNOWLEDGMENTS

\vspace{0.10in}

The authors would like to acknowledge useful conversations with Jerry
Stephenson, Tony Thomas, and Sid Coon.  This work was begun at the
INT Workshop:  ``Electromagnetic Interactions and the Few Nucleon
System''.

The authors thank INT for its hospitality and support.

\vspace{0.25in}

\noindent REFERENCES

\begin{enumerate}
\item For a review of the subject of CSB see G.\ A.\ Miller, B.\ M.\
K.\ Nefkens, and I.\ Slaus, Phys.\ Rep.\ {\bf 194} (1990) 1.\\

\item P.\ C.\ McNamee, M.\ D.\ Scadron, and S.\ A.\ Coon, Nucl.\
Phys.\ {\bf A249} (1975) 483, S.\ A.\ Coon, M.\ D.\ Scadron, and P.\
C.\ McNamee, Nucl.\ Phys.\ {\bf A287} (1977) 381; S.\ A.\ Coon and
R.\ C.\ Barrett, Phys.\ Rev.\ {\bf C36} (1987) 2189.\\

\item H.\ F.\ Jones and M.\ D.\ Scadron, Nucl.\ Phys.\ {\bf B155}
(1979) 409; M.\ D.\ Scadron, Phys.\ Rev.\ {\bf D29} (1984) 2076; S.\
A.\ Coon, B.\ H.\ J.\ McKellar, and M.\ D.\ Scadron, Phys.\ Rev.\
{\bf D34} (1986) 2784.\\

\item R.\ A.\ Brandenburg, S.\ A.\ Coon, and P.\ U.\ Sauer, Nucl.\
Phys.\ {\bf A294} (1978) 1752.\\

\item T.\ Goldman, J.\ A.\ Henderson, and A.\ W.\ Thomas, Los Alamos
Preprint LA-UR-91-2010.\\

\item  K.\ Maltman, Los Alamos Preprint LA-UR-92-2549.\\

\item J.\ Gasser and H.\ Leutwyler, Nucl.\ Phys.\ {\bf B250} (1985)
465.\\

\item J.\ F.\ Donoghue and B.\ R.\ Holstein, Phys.\ Rev.\ {\bf D40}
(1989) 2378; Phys.\ Rev.\ {\bf D40} (1989) 3700.\\

\item K.\ Maltman and D.\ Kotchan, Mod.\ Phys.\ Lett.\ {\bf A5}
(1990) 2457.\\

\item R.\ Dashen, Phys.\ Rev.\ {\bf 183} (1969) 1245.\\

\item R.\ H.\ Socolow, Phys.\ Rev.\ {\bf B137} (1965) 1221.\\

\item A.\ Manohar and H.\ Georgi, Nucl\ Phys.\ {\bf B234} (1984)
189.\\

\item A lucid discussion of the various possible chiral
transformation laws for fermion fields, and of the chiral quark model
may be found in:  H.\ Georgi, ``Weak Interactions and Modern Particle
Physics'' (Benjamin, Menlo Park, 1984).\\

\item S.\ Godfrey and N.\ Isgur, Phys.\ Rev.\ {\bf D32} (1985) 189.\\

\item S.\ Capstick and N.\ Isgur, Phys.\ Rev.\ {\bf D34} (1986)
2809.\\
\end{enumerate}

\pagebreak

\noindent {\bf TABLE 1.}  The slope of $\theta (q^{2})$ with $q^{2}$
as a function of $m, \delta m^{con}_{s}$

\vspace{0.10in}

\noindent \underline{\hspace{5.9in}}

\vspace{0.15in}
\begin{tabbing}
xxxxxxxx\=xxxxxxxxxxxxxxxxxxx\=xxxxxxxxxxxxxxxxxxxxxxxxxxx\=xxxxxxxxx
xxx\= \kill

\> {m(MeV)} \> {$\delta m^{con}_{s}$(MeV)} \> {$c_{1}(GeV^{-2})$} \\
\noindent \underline{\hspace{1.5in}} \>\underline{\hspace{1.8in}}
\>\underline{\hspace{2.5in}} \> \underline {\hspace{1.0in}} \\
\>\hspace{0.15in}202.5 \> \hspace{0.25in}175 \> .223 $\pm$ .026 \\
\> \> \hspace{0.25in}200 \> .234 $\pm$ .026 \\
\> \> \hspace{0.25in}225 \> .244 $\pm$ .029 \\
\> \hspace{0.17in}220 \> \hspace{0.25in}175 \> .211 $\pm$ .024 \\
\> \> \hspace{0.25in}200 \> .222 $\pm$ .025 \\
\> \> \hspace{0.25in}225 \> .231 $\pm$ .026
\end{tabbing}
\noindent \underline{\hspace{5.9in}}

\pagebreak

\noindent {\bf Figure Captions}\\

\noindent {\bf 1.} Pseudovector-coupling-induced contributions to the
meson propagator matrix.

\noindent {\bf 2.} Tadpole contributions to the meson propagator
matrix.

\end{document}